\documentclass[conference]{IEEEtran}
\IEEEoverridecommandlockouts

\usepackage{cite}
\usepackage{amsmath,amssymb,amsfonts}
\usepackage{graphicx}
\usepackage{iftex}
\ifPDFTeX\else
  \usepackage{fontspec}
\fi
\usepackage{textcomp}
\usepackage{xcolor}
\usepackage[hyphens]{url}
\usepackage{booktabs}
\usepackage{array}
\usepackage{multirow}
\usepackage{pifont}
\usepackage{tikz}
\usetikzlibrary{positioning,arrows.meta}

\usepackage[hidelinks]{hyperref}


\def\BibTeX{{\rm B\kern-.05em{\sc i\kern-.025em b}\kern-.08em
    T\kern-.1667em\lower.7ex\hbox{E}\kern-.125emX}}

\begin{document}

\ifdefined\pdfpagewidth
  \pdfpagewidth=8.5in
  \pdfpageheight=11in
\else
  
\fi

\pagenumbering{arabic}

\title{Exploring the Potential of Ultra-Low-Bit Inference on LLM Accelerators}

% Author list: compact block in the style of the group's other papers --
% numeric affiliation superscripts, affiliations on their own lines, and one
% email line at the end (no "Email:" label; same-domain addresses merged).
\author{
\IEEEauthorblockN{Hao Wang\textsuperscript{1},
Yang Shi\textsuperscript{1},
Jiageng Shi\textsuperscript{1},
Qiuran Li\textsuperscript{2}, and
Yaohua Wang\textsuperscript{1,*}}
\IEEEauthorblockA{1~~National University of Defense Technology\\
Key Laboratory of Advanced Microprocessor Chips and Systems, Changsha, China\\
2~~Zhejiang University\\
\{hao\_wang, shiyang14, shijiagengsjg, yhwang\}@nudt.edu.cn, qiuranli@zju.edu.cn\\
*~~Corresponding author}
}

% Explicit PDF metadata: hyperref cannot derive it reliably from the
% IEEEtran author-block macros.
\hypersetup{
  pdftitle={Exploring the Potential of Ultra-Low-Bit Inference on LLM Accelerators},
  pdfauthor={Hao Wang, Yang Shi, Jiageng Shi, Qiuran Li, Yaohua Wang},
  pdfkeywords={vector quantization, large language model, dequantization, systolic array}
}

\maketitle
\thispagestyle{plain}
\pagestyle{plain}

\begin{abstract}
Vector quantization (VQ) enables ultra-low-bit compression of large language models (LLMs) by encoding groups of values as compact codebook indexes. However, reconstructing these values for computation requires frequent codebook lookups, which conflict with the dataflows of conventional LLM accelerators and limit the performance gains from VQ compression. Consequently, prior work has resorted to VQ-specific accelerator designs, which either incur substantial additional hardware overhead or support only a limited range of inference scenarios.
In this work, we argue that existing systolic-array-based LLM accelerators already possess the essential hardware resources for efficient VQ inference, but lack the mechanisms to effectively exploit them.
To bridge this gap, we present VQACC, an architectural enhancement that
jointly redesigns the existing operand buffer's access logic and data mapping
to reuse its storage and port resources for codebook lookups, seamlessly
integrating dequantization into the normal operand-addressing dataflow. Evaluation shows that VQACC achieves a
2.63$\times$ speedup over the EVA baseline, with additional hardware
occupying only 0.17\% of the compute array's area.
% [reported] VQ accuracy/compression motivation: VQ-LLM, Abstract and Sec. I/II, Fig. 2.
% [reported] Separately synthesized shift + index-address logic / SA area:
% (2747.135941 + 1037.568003) / 2206252.600816 = 0.17154%; notes/area-comparison.json.
% Shift network = 32-lane rotator (rtl_fp16/evaluation/20261118_rotate32, 2026-11-18).
% This is a component-area ratio, not total added hardware or post-layout area.
% [modeled] evaluation/data/plot_values.csv, overall_average:
% mean(T_EVA/T_FP16) / mean(T_VQACC/T_FP16)
% = 1.464630042086266 / 0.48556093266725064 = 3.016367.
% 288 workloads: 3 LLMs x 3 joint W/KV configurations x batches 1--32, KV=2K.
% EVA-OL32 uses FP16 KV and a 32x8 construction formula; VQACC uses
% simplified CQ traffic (16 KiB/head/operand), excluding KV encoding.
% This is the current model comparison, not measured or equal-area speedup.
\end{abstract}


% Body text, figures, and evidence comments are inherited from the DAC 2027
% workspace at ../vqacc_dac_2027, where the evaluation data and provenance
% live. This directory carries only the arXiv manuscript; the [reported] /
% [modeled] / [design] tags in the comments refer to that workspace's notes.
\section{Introduction}
\label{sec:intro}

% Reserve the current column for the abstract; float Fig. 1 atop the next.
\suppressfloats[t]
\begin{figure}[t]
\centering
\includegraphics[width=\columnwidth]{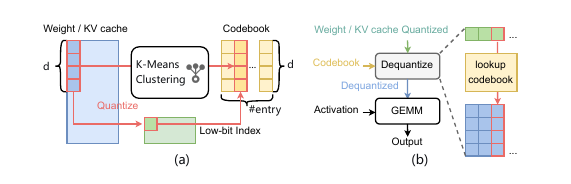}
\caption{(a) The principle of VQ. (b) VQ dequantization flow.}
\label{fig:vq}
\end{figure}

The success of large language models (LLMs) in language understanding and
generation places increasing demands on hardware memory capacity and
bandwidth~\cite{gmlake2024,sglang2024,vllm2023,veltair2022}.
LLM inference stores large model weights and key--value (KV) caches that
retain information from previously processed tokens. KV caches grow with
context length and the number of concurrent requests, and both weights and KV caches
must be accessed repeatedly during token generation. Quantization represents
these data with fewer bits, reducing transfers and allowing larger models and
longer contexts within a given memory
budget~\cite{vqllm2025,gptq2022,squant2022,olive2023,ant2022}.

% [reported] VQ-LLM Sec. I and Sec. II-A/B, Fig. 2: vector compression,
% cross-element information, and improved accuracy/compression tradeoffs.
Vector quantization (VQ) extends this opportunity to ultra-low-bit inference
with 4-bit or lower precision per value. Unlike commonly used element-wise quantization, which
compresses each value independently,
VQ jointly represents a group of values as a vector. It learns a shared
codebook of representative vectors and replaces each original vector with
the low-bit index of its closest representative (Fig.~\ref{fig:vq}(a)).
The index thus encodes the pattern across multiple elements, exploiting their
correlations to preserve information that independent quantization loses.
For LLMs, VQ can achieve higher accuracy than element-wise quantization at
an effective precision of 4-bit per value, or retain comparable accuracy at 2-bit~\cite{aqlm2024,gptvq2024,cq2024}, and some
methods even achieve 1-bit compression~\cite{aqlmrelease}.
Element-wise quantization, in contrast, is commonly limited to 4-bit by
substantial accuracy loss at lower bit widths~\cite{awq2024,kvquant2024,qserve2025}.
This improved accuracy--compression tradeoff makes VQ attractive for
ultra-low-bit LLM inference.
% [design] Ultra-low-bit denotes nominal precision <=4 bits per value here.
% [reported] 4-bit/2-bit accuracy comparison: VQ-LLM Sec. I and Fig. 2.
% [reported] 1-bit weights: AQLM official release, April 2025, Llama-2-7b
% 1x8g8 (nominal 1-bit). VQ-LLM reference [69] concerns KV caches, not weights.

However, a gap remains between VQ's strong compression capability and its
actual performance gains. As Fig.~\ref{fig:vq}(b) shows, VQ introduces
dequantization before LLM matrix computations, retrieving vectors from
codebooks using their indexes. Commercial accelerators handle these frequent
lookups poorly, which erodes the expected speedup from VQ: some VQ operators
even run slower than their FP16 counterparts~\cite{vqllm2025,eva2026}. Specialized accelerators for VQ-based LLMs
address this problem at the cost of additional hardware or limited applicability.
Vortex builds a custom dequantization pipeline with units such as additional
codebook buffers and lookup staging buffers, which incurs expensive hardware
overhead~\cite{vortex2026}.
EVA bypasses conventional dequantization through a mathematical transformation,
but supports VQ only for model weights, with benefits concentrated in
small-batch LLM decoding~\cite{eva2026}.
% [reported] VQ-LLM Sec. VII-D, Fig. 16: existing VQ kernels can exceed FP16 latency.
% [reported] Vortex Sec. 3 / Fig. 3: LUU and extended VPU; Sec. 4.1:
% lookup and accumulation flows; Sec. 5 / Table 4: 128 KiB double-buffered
% LUU storage.
% [reported] EVA Sec. III (GEMV-like workloads), Sec. IV-C (batch scaling).

This work starts from a key observation: in existing LLM accelerators based
on systolic arrays (SAs), operand-buffer addressing is naturally compatible
with codebook lookup, and its capacity and ports can also support storing
multiple codebooks of typical size and performing parallel, conflict-free lookups.
What is missing is a mechanism to exploit these resources effectively.
We therefore propose VQACC, a VQ-enhanced architecture built on existing
SA-based LLM accelerators. VQACC jointly redesigns operand-buffer access
logic and data mapping to integrate dequantization into operand addressing
with minimal hardware overhead. Four complementary designs realize this
approach. First, we propose a mapping strategy that reuses the existing operand
buffer to store codebooks and the indexes representing quantized values.
Second, building on this mapping, we reuse existing buffer ports to read
indexes and design an index address generator to convert them into buffer
addresses. Reads through the existing ports at these addresses then directly
return the codebook lookup results. Third, we design a lightweight shift
layout unit that realigns lookup results on the fly into the operand order
required by the SA, without additional latency. Finally, we refine the existing dataflow schedule at tile
granularity to reduce codebook-related memory traffic.

Evaluations across multiple LLMs, VQ algorithms, and inference scenarios show
that VQACC achieves speedups of 2.63$\times$ and 1.10$\times$ over EVA and
Vortex, respectively, while its added logic occupies only 0.17\% of the
compute array's area, 125.0$\times$ less than Vortex's dedicated
dequantization hardware.

\section{Background and Related Work}
\label{sec:bg}

\begin{table}[!t]
\centering
\caption{PARAMETERS OF VQ ALGORITHMS}
\label{tab:vq-notation}
\setlength{\tabcolsep}{5pt}
\footnotesize
\begin{tabular}{@{}llc@{}}
\toprule
Item & Description & Example values \\
\midrule
$d$ & Number of values per vector & 2/4/8 \\
$\#\mathrm{entry}$ & Number of entries in a codebook & 256 \\\bottomrule
\end{tabular}
\end{table}

% [reported] notes/background-vq-accuracy.json: GPTVQ-2 dPPL 1.59 (VPTQ Table 4),
% AWQ-2 ~4.3 (VQ-LLM Fig. 2 reading), AQLM-1 2.73 (provisional, AQLM release),
% CQ-2/KVQuant-2 0.30/0.38 and CQ-4/KVQuant-4 0.02/0.06 (CQ Table 1).
% Mixed-source literature overview, not a controlled accuracy evaluation.
% [reported] Panel (c): notes/a100-cq2-summary.json; A800, FP16 30.720 us,
% CQ-2 61.440 us; B=1, H=32, D=128, T=1024, synthetic CQ-format input.
% Declared here, while Sec. II starts in the left column of page 2, so that
% the figure lands atop the right column of the same page.
\suppressfloats[t]
\begin{figure}[t]
\centering
\includegraphics[width=\columnwidth]{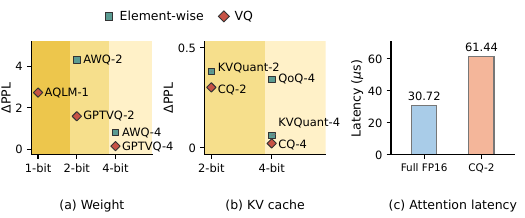}
\caption{(a,b) Accuracy comparison of VQ and element-wise quantization
for (a) weights and (b) KV caches~\cite{vqllm2025,vptq2024,cq2024,aqlmrelease,aqlm2024,gptvq2024,awq2024,kvquant2024,qserve2025},
where $\Delta$PPL is the perplexity increase over FP16 (lower is better).
(c) Attention latency comparison between Full FP16 and CQ-2 for batch-1 decoding
of Llama-2-7B~\cite{llama2} on A800 with a 1024-token KV cache.}
\label{fig:vq-accuracy}
\end{figure}

\subsection{Quantization in LLM Inference}
\label{sec:bg-quant}

LLM inference consists of a prefill phase and a decoding phase. Prefill
processes the entire prompt at once, whereas decoding generates one token at
a time. Weights and KV caches are
fetched repeatedly for each generated token with little reuse, so decoding is
typically bound by memory bandwidth. Batching multiple requests amortizes
weight fetches across the batch, but each request keeps its own KV cache, so
KV traffic still grows with both the batch size and the context length. KV
caches also occupy memory capacity in proportion to these two factors,
limiting the number of concurrent requests and thus the benefit of
batching~\cite{sarathi2024,bytedance2026}. Memory bandwidth therefore remains a severe
challenge for LLM decoding.
% [reported] Phrasing follows EVA Sec. I and Sec. II-A: decoding generates one
% token at a time and fetches weights per generated token without reuse.

Quantization stores weights and KV caches in fewer bits, reducing both memory
traffic and footprint and thereby delivering significant speedups in LLM
decoding~\cite{deepseekv3_2025,omniquant2024}. The same quantized weights usually serve prefill as well, because
keeping a separate high-precision copy for prefill would offset the memory
savings, which memory-constrained inference systems cannot afford~\cite{atom2024,ondevice2025}.

\subsection{Vector Quantization}
\label{sec:bg-vq}

Table~\ref{tab:vq-notation} summarizes the two parameters that characterize
a VQ algorithm. VQ encodes each $d$-value vector with a
$\log_2(\#\mathrm{entry})$-bit index into a codebook of $\#\mathrm{entry}$
representative vectors, giving an effective precision of
$\log_2(\#\mathrm{entry})/d$ bits per value. Each codebook serves one scope of
a tensor, such as a whole weight matrix, a weight block, or a KV
head~\cite{vqllm2025}.
% [reported] VQ-LLM Sec. II-A/Tbl. I (vector size, #Entry) and Sec. V-C:
% "algorithms train different codebooks for different parts of a tensor".
Like prior VQ LLM accelerators~\cite{eva2026,vortex2026}, VQACC targets
256-entry codebooks, the practical configuration that mainstream VQ
algorithms typically offer~\cite{aqlm2024,cq2024}. Such a codebook balances accuracy
against on-chip storage: a single FP16 codebook occupies at most 4\,KiB for
$d\le8$, and each index fits in exactly one byte. With $d=2$, 4, and 8, it
yields a precision of 4-bit, 2-bit, and 1-bit per value, respectively.
% [reported] EVA Sec. VI design-space exploration: "2^n = 256 (n = 8) provides
% the most practical trade-off between compute efficiency and memory cost".
% Vortex Sec. 4.3: "we choose b = 8 as a practical design point".
% AQLM 1x8/2x8 and CQ c8b configurations use 8-bit (256-entry) codebooks.
% [design] 256 x 8 x 2 B = 4 KiB.

As Fig.~\ref{fig:vq-accuracy}(a,b) shows, VQ preserves accuracy better than
element-wise quantization at the same bit width for both weights and KV
caches, and its advantage grows as the bit width decreases, giving VQ a clear
accuracy--compression advantage at 4-bit and below. This advantage, however,
comes at the cost of expensive dequantization: on existing commercial
accelerators, hardware and software stacks designed for regular LLM dataflows
handle its frequent codebook lookups poorly, causing heavy
on-chip data movement and bank conflicts, so VQ can even run slower than FP16
(Fig.~\ref{fig:vq-accuracy}(c)).

\subsection{VQ-Based LLM Accelerators}
\label{sec:bg-acc}

Several accelerators customized for VQ-based LLMs have been proposed to
overcome the inefficiency of dequantization, but they either incur
substantial hardware overhead or support only a limited range of inference
scenarios.

Vortex introduces dedicated hardware for VQ dequantization~\cite{vortex2026}.
An additional multi-bank codebook buffer holds multiple codebook copies to
support parallel, conflict-free lookups, and an additional lookup staging
buffer reorders and aligns the lookup results before they enter the compute
array.
These additional buffers, however, incur expensive hardware overhead.

EVA instead avoids weight dequantization by moving the lookup to the output
side~\cite{eva2026}. It multiplies the activations with the weight codebook,
treats the products held in the output buffer as a table of partial sums, and
reuses the multiple banks of the output buffer to select and accumulate table
entries in parallel according to the weight indexes. Because both the table
and the number of lookups grow with the batch size, EVA's benefits
concentrate in small-batch decoding. Moreover, the transformation depends on
how a weight tensor is partitioned into blocks that share one codebook, which
restricts it to weight VQ.

In contrast to these designs, our approach adds VQ support to existing LLM
accelerators at minimal hardware cost, and it applies efficiently to all LLM
quantized inference scenarios.

\section{VQACC Design}
\label{sec:design}

\subsection{Overview}
\label{sec:design-overview}

\begin{figure}[t]
\centering
\includegraphics[width=\columnwidth]{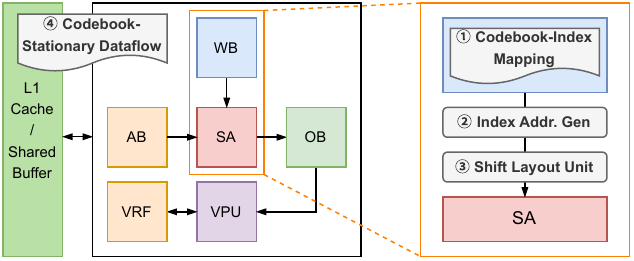}
\caption{Overview of the VQACC architecture.}
\label{fig:overview}
\end{figure}

In this section, we present VQACC, an architectural enhancement that
accelerates VQ dequantization by efficiently reusing the resources that an LLM
accelerator already provides. Fig.~\ref{fig:overview} shows the overall
architecture of VQACC. VQACC builds on typical SA-based accelerator
architectures~\cite{gemmini2021,eyerissv2_2019,maestro2018}: the SA reads activations from the activation
buffer (AB) and weights or KV caches from the weight buffer (WB), and writes
results to the output buffer (OB); a vector processing unit (VPU) with its
register file (VRF) executes operators other than matrix multiplication; and
an L1 cache or shared
buffer passes VPU results to the SA on chip, as kernels such as
FlashAttention require~\cite{flashattention2022,vllm2023}. AB, WB, and OB are built from
multi-bank SRAMs, each bank with two read/write ports, which supports
ping-pong scheduling between external data transfers and operand supply to the
SA. On this basis, VQACC enhances only the operand path from WB to the SA so
that it can additionally support the VQ dequantization flow. Four designs
realize this change, numbered \ding{172}--\ding{175} in
Fig.~\ref{fig:overview} and detailed in
Sec.~\ref{sec:design-mapping}--\ref{sec:design-dataflow}.

\subsection{Codebook-Index Mapping}
\label{sec:design-mapping}

\begin{figure}[t]
\centering
\includegraphics[width=\columnwidth]{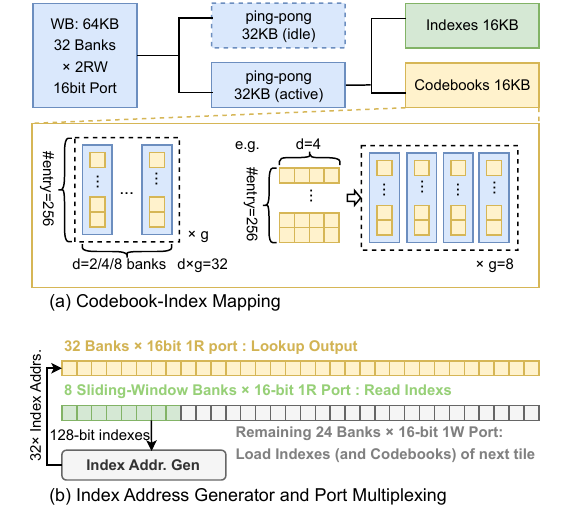}
\caption{(a) Codebook-index mapping. (b) Index address generator and port
multiplexing.}
\label{fig:placement}
\end{figure}

First, we propose a codebook-index mapping that stores codebooks and indexes
in WB's existing banks in a layout that enables efficient lookups. We assume
that WB holds 64\,KiB in 32 dual-port banks of 16-bit width for FP16 operands.
As Fig.~\ref{fig:placement}(a) shows, VQACC splits each 32\,KiB ping-pong
half of WB into 16\,KiB of indexes and 16\,KiB of codebooks. While the SA
consumes the active half, the idle half receives the indexes and codebooks of
the next tile. Based on the target codebook configuration introduced in
Sec.~\ref{sec:bg-vq}, both regions meet their storage requirements with the
capacity WB already has. For the index region, 16\,KiB of indexes can decode into
64\,KiB of FP16 values even at $d=2$, twice the 32\,KiB
capacity limit of an unquantized half. Any unquantized operand tile that fits
in a half therefore still fits in the index region after VQ. For the codebook
region, VQACC maps $32/d$ codebooks (copies of the same codebook or different
codebooks), which exactly fill the 16\,KiB codebook region, as a 256-entry
FP16 codebook occupies $512d$\,B. For the placement of these codebooks, we
specify that each codebook is distributed across $d$ adjacent banks, with each
bank holding one element position, so a single access to these $d$ banks
retrieves a complete vector without bank conflicts. The 32 banks thus form
$32/d$ groups that can look up in parallel, fully using all 32 banks every
cycle.
% [design] 256 x d x 2 B = 512d B; (32/d) x 512d B = 16 KiB for every d.
% 16384 indexes x d values x 2 B = 32d KiB >= 64 KiB at d=2.

\subsection{Index Address Generator and Port Multiplexing}
\label{sec:design-addr}

Second, we use an index address generator together with a port multiplexing
strategy that provides, at the same time, the bandwidth a lookup needs for its
index input and for its output.
VQACC first provides one port of each of the 32 banks as a read port that
outputs codebook lookup results. Among
the remaining ports, we choose eight adjacent banks to provide read ports that
fetch $8\times16$ bits, i.e., 128 bits of indexes as the input to the
codebook lookup. Because a 128-bit index word holds sixteen 8-bit indexes, one
word covers all $32/d$ codebooks even at $d=2$, so the indexes read in one
cycle guarantee that every codebook is supplied at least once. These eight
adjacent banks are therefore defined as an index window: we divide the 32
banks into four windows and use a 4:1 multiplexer to move the window across the
32 banks. We add an index address generator that turns each 128-bit index word
into 32 index addresses, one per bank, and drives the lookup read ports of the
banks in the next cycle. Finally, the remaining 24 banks provide write ports
that write the indexes of the next tile into the idle half, along with new
codebooks when the codebooks need to be replaced.
% [design] rtl_index_addr_gen: registered 128-bit word; 16/8/4 indexes per
% cycle for d=2/4/8 (word lasts 1/2/4 cycles); 2-bit prefix + 8-bit index.

\begin{figure*}[t]
\centering
\includegraphics[width=\textwidth]{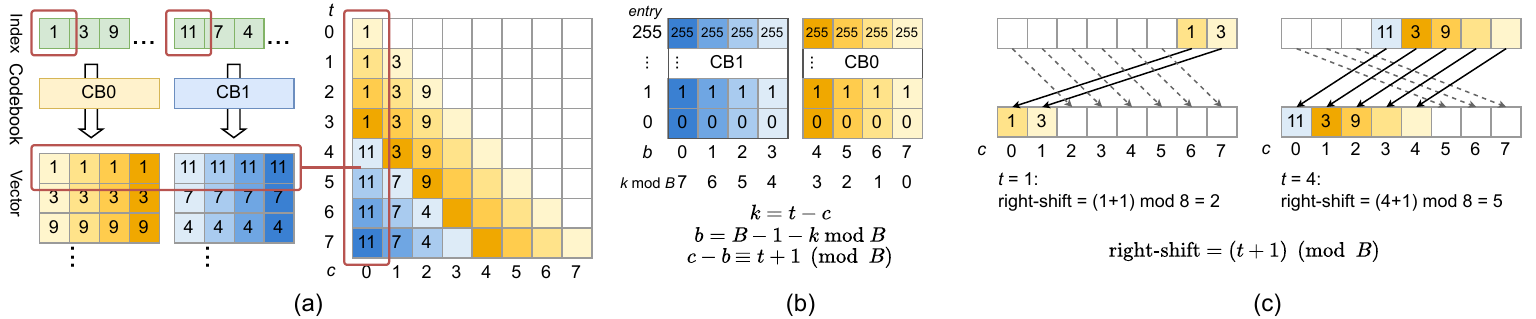}
\caption{Operation of the shift layout unit, shown for a WB with $B=8$ banks
holding two $d=4$ codebooks (CB0 and CB1) and an output-stationary SA with 8
columns. (a) Operand layout required by the SA. (b) Reverse codebook
placement. (c) In cycles $t=1$ and 4, the bank outputs (top) are cyclically
right-shifted by $(t+1)\bmod B$ onto the correct SA columns (bottom).}
\label{fig:rotate}
\end{figure*}

This port multiplexing sustains the full operand bandwidth of the SA with
little index traffic. The read ports of the 32 banks return 32 FP16 values
per cycle, exactly the number of operands the SA consumes, so the SA is fed
at the same rate as with unquantized FP16 operands. Meanwhile, each cycle
consumes only $32/d$ bytes of indexes, $2d\times$ less on-chip traffic than
FP16 weights. Excluding codebook loads, the write ports of the 24 loading
banks provide 48\,B/cycle, which easily keeps up with this index consumption
under ping-pong scheduling, and the compression benefit extends to on-chip
data movement.
% [design] 24 banks x 16-bit write port = 48 B/cycle >= 32/d B/cycle (<= 16).

\subsection{Shift Layout Unit}
\label{sec:design-shift}

Third, we design a lightweight shift layout unit that converts WB's lookup
results on the fly into the operand layout the SA consumes. The unit works
together with the output-stationary SA; we explain it with the example in
Fig.~\ref{fig:rotate}, which uses $B=8$ WB banks for clarity and extends
directly to $B=32$. As Fig.~\ref{fig:rotate}(a) shows, each SA column receives
one operand per cycle and starts one cycle after its left neighbor, forming a
diagonal wavefront: in cycle $t$, column $c$ consumes the operand at reduction
position $k=t-c$. Because VQ vectors usually lie along the
reduction dimension, the elements of looked-up vectors must enter the same SA
column over consecutive cycles. To enable parallel, conflict-free lookups, however, the $d$
elements of each codebook vector are spread over $d$ separate banks, so a
lookup returns them side by side, and the results must be transposed before
they can flow down the SA columns (red boxes in Fig.~\ref{fig:rotate}(a)).

To solve this efficiently, we design a shift-based transpose built on reverse
placement, which folds the layout alignment into the SA's operand supply.
First, VQACC spreads the reduction dimension across all $B$ banks in reverse
order: from the SA's view, the operand element at reduction position $k$ is
stored in bank $b=B-1-(k\bmod B)$ (Fig.~\ref{fig:rotate}(b)). Within a codebook, each
vector is stored in reverse, e.g., elements 0--3 of CB0 in banks 7--4. Some VQ
algorithms, such as CQ for key cache~\cite{cq2024}, use different codebooks for consecutive
vectors along the reduction dimension, so one reduction span is restored from
several codebooks; the order across codebooks is therefore reversed as well,
as CB0 occupies banks 7--4 and CB1 banks 3--0 in the example.
Next, we derive which bank feeds which SA column. In cycle $t$, column $c$ needs
position $k=t-c$, which reverse placement puts in bank $b\equiv B-1-t+c
\pmod{B}$, so the distance from this target bank to the column is
$c-b\equiv t+1 \pmod{B}$, independent of the column. Each bank therefore reads its data at the index addresses
that produce the diagonal wavefront of Fig.~\ref{fig:rotate}(a), supplied by
the index address generator (Sec.~\ref{sec:design-addr}), and the unit right-shifts all bank outputs by
$(t+1)\bmod B$ (a standard $B$-lane, 16-bit rotator suffices), delivering
every operand to its correct column at once. As
Fig.~\ref{fig:rotate}(c) shows, right shifts by 2 and 5 in cycles $t=1$ and 4
exactly reproduce these two rows of Fig.~\ref{fig:rotate}(a). For the special VQ algorithms that lay
codebook vectors along the output dimension (e.g., CQ for the value cache~\cite{cq2024}), no
transpose is needed; the vectors are stored in natural order with a zero shift.
The unit integrates seamlessly
into the existing datapath without adding any cycle or pipeline stage, so from
the SA's view the operand supply is identical to the normal dataflow.
% [design] column c takes bank (c-r) mod B, r = (t+1) mod B (vectors along K)
% or 0 (vectors along N); bank b serves column (b+r) mod B. Cycle-level check:
% rtl_fp16/tests/model_rotate32_dataflow.py. RTL rtl_fp16/rtl/vq_rotate32.sv;
% Xcelium PASS, 265,216 checks.
% [reported] rtl_fp16/evaluation/20261118_rotate32/.../result.json (TSMC28 HPC+
% 12T SS 0.9 V 125 C, DC O-2018.06-SP1, compile_ultra -retime, 1 ns virtual
% clock, 0.1 ns uncertainty, 0.2/0.2 ns I/O): 2747.135941 um^2, combinational,
% setup slack +0.003826 ns. vq_full_shuffle 32x32 crossbar 10122.335708 um^2;
% 1 - 2747.136/10122.336 = 72.86%.
% Open [design]: the synthesized index address generator broadcasts one index
% to the d banks of a group; the per-bank routing (bank b gets the address of
% column (b+t+1) mod B) is not in the synthesized RTL or the 3,785 um^2.

\subsection{Codebook-Stationary Dataflow}
\label{sec:design-dataflow}

\begin{figure}[t]
\centering
\includegraphics[width=\columnwidth]{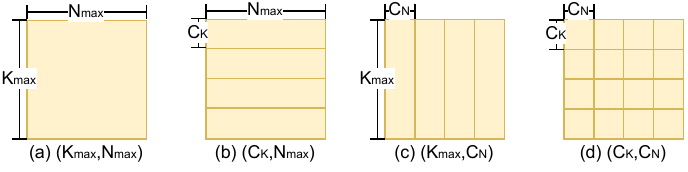}
\caption{Codebook-sharing partitions of a $K_{\max}\times N_{\max}$ operand.
For an operand dimension $D\in\{K,N\}$, let $C_D$ be the size of the region
that shares one codebook along $D$, and $D_{\max}$ means that this dimension is
not partitioned.}
\label{fig:partition}
\end{figure}

Finally, we design a codebook-stationary dataflow at tile granularity that
keeps each codebook resident in WB for as long as possible, so that codebook
loads cost less. As Fig.~\ref{fig:partition} shows, VQ algorithms often
partition an operand into regions that each use their own codebook. A
conventional dataflow schedules tiles without regard to these regions, so
a single tile frequently crosses codebook boundaries; every crossing reloads
a codebook, consuming DRAM and L1 bandwidth and stalling the lookups that
depend on it. Let $T_D$ be the tile size along $D$ in the conventional
dataflow and $T'_D$ the tile size that VQACC uses instead. When $C_D>B$ (bank number of WB),
VQACC sets $T'_D=\min(T_D,C_D)$ so that no tile straddles two codebook
regions. When $C_D\le B$, VQACC sets $T'_D=nC_D$ with the largest positive
integer $n$ satisfying $nC_D\le B$, as the codebook mapping of
Sec.~\ref{sec:design-mapping} can hold and decode different codebooks
simultaneously; a tile thus grows as large as the lookup bandwidth allows.
Furthermore, while keeping the operand and result residency of the
conventional dataflow, VQACC reorders tiles to traverse the region of the
current codebook first, so that the codebook stays resident in WB across
many tiles.
% [design] Table~II lists C_K/C_N per configuration. GPTVQ: T'_K = min(512,256)
% = 256, T'_N = 32; a 256x256 block = eight 256x32 tiles that reuse the A tile
% and codebook (8x fewer codebook loads than a per-output-tile order).
% CQ: T'_K (keys) / T'_N (values) = 32, i.e. 32/d codebooks per 16 KiB region.
% Keys: the model's K32xN512 tile equals sixteen K32xN32 tiles run back to back
% on the resident slice (SA-bound once the codebook stays in WB), per author.
% exp/reuse_schedule.py make_plan()/tiles(); exp/check_cq_slices.py.

\section{Evaluation}
\label{sec:eval}

\subsection{Experimental Setup}
\label{sec:eval-setup}
% [modeled] All latency numbers come from exp/two_lane.py and its schedules;
% paper data: evaluation/data/*.csv, provenance.json.

% Keep both setup tables together atop the next column, after their first
% reference instead of above the end of Sec. III.
\suppressfloats[t]
\begin{table}[t]
\centering
\caption{VQ CONFIGURATIONS IN EVALUATION}
\label{tab:eval-vq}
\setlength{\tabcolsep}{3pt}
\footnotesize
\begin{tabular}{@{}llrrcc@{}}
\toprule
Scheme & Target & $d$ & Bits & Codebook partition & Abbr. \\
\midrule
AQLM-1 & Weight & 8 & 1 & $(K_{\max},N_{\max})$ & W1 \\
GPTVQ-2 & Weight & 4 & 2 & $(C_K{=}256,C_N{=}256)$ & W2 \\
CQ-2 & \begin{tabular}[c]{@{}l@{}}Key /\\Value\end{tabular} & 4 & 2 & \begin{tabular}[c]{@{}c@{}}$(C_K{=}4,N_{\max})$/\\$(K_{\max},C_N{=}4)$\end{tabular} & KV2 \\
CQ-4 & \begin{tabular}[c]{@{}l@{}}Key /\\Value\end{tabular} & 2 & 4 & \begin{tabular}[c]{@{}c@{}}$(C_K{=}2,N_{\max})$/\\$(K_{\max},C_N{=}2)$\end{tabular} & KV4 \\
\bottomrule
\end{tabular}
\end{table}

\begin{table}[t]
\centering
\caption{HARDWARE CONFIGURATION}
\label{tab:eval-architecture}
\setlength{\tabcolsep}{5pt}
\renewcommand{\arraystretch}{1.12}
\footnotesize
\begin{tabular}{@{}l|l@{}}
\toprule
\textbf{SA} & $32\times32$ FP16, output stationary, 1\,GHz \\
\midrule
\textbf{AB/WB/OB} & 64\,KiB each, 1\,GHz \\
\midrule
\textbf{VRF\&VPU} & 64\,KiB, 32 lanes, 1\,GHz \\
\midrule
\textbf{L1 Cache} & 64\,B/cycle, 1\,GHz \\
\midrule
\textbf{DRAM} & 25.6\,GB/s (DDR4-3200) \\
\bottomrule
\end{tabular}
\end{table}

We evaluate Llama-2-7B~\cite{llama2}, Qwen3-8B~\cite{qwen3modelcard}, and
Mistral-7B~\cite{mistral7b} under
the VQ configurations in Table~\ref{tab:eval-vq}, using \textbf{AQLM-1 (W1)}~\cite{aqlm2024} and
\textbf{GPTVQ-2 (W2)}~\cite{gptvq2024} for weights and \textbf{CQ-2 (KV2)} and
\textbf{CQ-4 (KV4)}~\cite{cq2024} for KV caches. The overall comparison
combines them into W1KV2, W2KV2, and W2KV4 and covers decoding with batch
sizes 1--32 and a 2K-token KV cache.

Building on SCALE-Sim~\cite{scalesim2018,scalesim2025}, we develop a cycle-level simulator
for VQ LLM inference, with the architecture parameters in
Table~\ref{tab:eval-architecture}. We compare
VQACC with four designs built on the same accelerator configuration. \textbf{Full
FP16} uses full FP16 operands. \textbf{VPU Dequant} dequantizes in the VPU;
we extend the VPU with VQ-specific instructions so that it performs
conflict-free lookups from the VRF at the same lookup bandwidth as VQACC. Its
gap therefore lies mainly in the extra on-chip traffic of writing the
reconstructed FP16 operands back to WB through L1. \textbf{EVA-OL32} follows
EVA~\cite{eva2026} and reuses 32 output-buffer banks for output lookup; it
keeps KV caches in FP16 and does not support prefill. \textbf{Vortex-IL}
uses the dedicated dequantization pipeline of Vortex~\cite{vortex2026} to look
up operands on the input side, with additional $2\times16$\,KiB ping-pong
codebook buffers and $2\times8$\,KiB ping-pong lookup staging buffers. Vortex
additionally integrates an EVA-style mechanism that is compatible with this
pipeline and selects between the two according to which suits the scenario;
since EVA is already part of our comparison and VQACC can incorporate EVA
equally cheaply, we do not evaluate that mechanism separately. Moreover, we
do not perform online VQ (clustering) of the dynamically generated KV caches on
any of these designs; as in common VQ deployments, it is offloaded to an off-chip
processor~\cite{pqcache2025} and does not affect LLM inference. We implement
VQACC and the baseline designs in SystemVerilog and synthesize them in
28\,nm at 1\,GHz to compare area. SRAM buffer areas come from a commercial
memory compiler and SRAM access energy from CACTI~\cite{cacti7}; DRAM energy
uses the DDR4 access energy reported in GANAX~\cite{ganax2018}.
% [modeled] Self-written tile-level model (exp/two_lane.py, reuse_schedule.py,
% b3_schedule.py); SA microtile K+M+N-2 per SCALE-Sim OS
% (docs/discussion/scalesim-os-study.md); no SCALE-Sim code is used.
% Excludes KV-cache encoding; non-GEMM ops = DRAM traffic; FP16 operands,
% codebooks and psums; prefill on quantized weights with FP16 KV.
% CQ: one 16 KiB codebook slice (32 channels) per K/V tile; QK K32xN512, PV
% K512xN32 for VPU Dequant., Vortex-IL and VQACC (exp/check_cq_slices.py).
% VPU lookup 64 B/cycle = VQACC's 32 FP16/cycle (reuse_schedule vpu_bpc).
% RTL: rtl_fp16 (SA, shift network), rtl_index_addr_gen, rtl_eva_ol32; no
% RTL for VPU Dequant. or Vortex-IL (its lookup logic is charged as VQACC's).
% Memory compiler: area only (notes/area-comparison.json); energy uses CACTI 7
% and 15 pJ/bit DRAM (notes/energy-model.json); DRAMsim3 is not used.

\subsection{Decoding Performance}
\label{sec:eval-overall}
\begin{figure}[!t]
\centering
\includegraphics[width=\columnwidth]{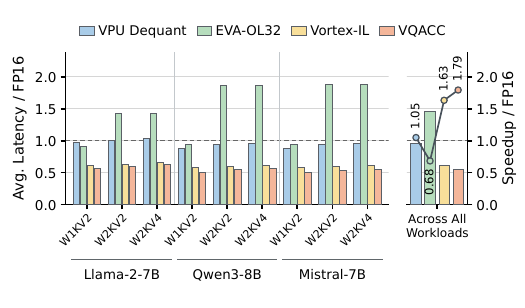}
\caption{Decoding latency normalized to Full FP16, averaged over batch
sizes 1--32 with a 2K-token KV cache.}
\label{fig:eval-overall}
\end{figure}
% [modeled] evaluation/data/plot_values.csv, overall_average (288 workloads):
% normalized latency FP16 1, VPU 0.9532, EVA 1.4640, Vortex 0.6124, VQACC 0.5575.
% Per-workload T_Vortex/T_VQACC = 1.038--1.165; each 8 KiB Vortex-IL subtile
% is costed as its own SA task (docs/experiment/b4-vortex.md).
% evaluation/data/decode.csv: per-workload T_FP16/T_VQACC = 1.429--2.327;
% batch-mean T_FP16/T_VQACC = 2.20 at b1, 1.61 at b32.
% batch-mean T_EVA/T_FP16 = 0.152 at b1; batch-mean T_EVA/T_VQACC = 0.95 at
% b4 and 1.20 at b5; EVA faster than VQACC only at b<=6 (33/288 workloads);
% batch-mean T_EVA/T_FP16 = 0.803 at b8 and 1.011 at b9.
% Qwen3-8B W2KV2 b32, one layer: attention (QK+PV) is 46% of VQACC latency;
% T_FP16/T_VQACC = 1.40 for attention vs 1.90 for linear layers.

Across all models, VQ algorithms, and decoding configurations, VQACC is the
fastest design on average (Fig.~\ref{fig:eval-overall}). Relative to Full FP16,
VPU Dequant, EVA-OL32, and Vortex-IL, it delivers average speedups of
1.79$\times$, 1.71$\times$, 2.63$\times$, and 1.10$\times$, respectively. That
Full FP16 and VPU Dequant differ by only a few percent shows that optimizing
the lookup operation alone cannot solve the VQ dequantization bottleneck: once
VQ compression removes most of the DRAM accesses, overall performance becomes
limited by on-chip execution time, so the additional on-chip traffic produced
by dequantization is an equally significant cost. Although Vortex-IL builds a
dedicated dequantization pipeline that optimizes both the lookup operation and
the on-chip data traffic, it still leaves performance headroom compared with
VQACC. The reason is that
Vortex-IL relies on lookup staging buffers to pass the dequantized operands to
the SA, and the capacity of these buffers limits the largest tile the SA can
execute at once. By contrast, VQACC keeps dequantization entirely within the
existing data path, so in some scenarios it can use larger tiles and obtain
higher SA utilization by draining the array less often. EVA-OL32
transforms the dequantization flow itself, which makes it the only design
whose performance degrades sharply as the batch grows and which cannot
support VQ inference for the KV cache; over the full multi-batch decoding
range, it can even fall below Full FP16.

\begin{figure}[!t]
\centering
\includegraphics[width=0.95\columnwidth]{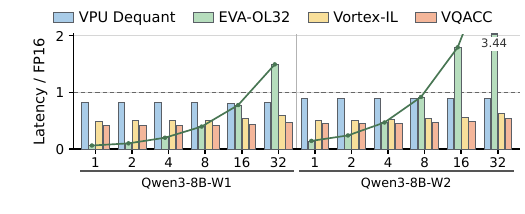}
\caption{Decoding latency of one Qwen3-8B feed-forward network (FFN) block with batch sizes
1--32, normalized to Full FP16.}
\label{fig:eval-batch}
\end{figure}
\begin{figure}[!t]
\centering
\includegraphics[width=0.95\columnwidth]{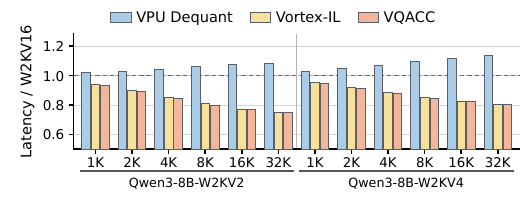}
\caption{Decoding latency of Qwen3-8B with CQ-2/CQ-4 KV caches at batch 16
and context lengths 1K--32K, normalized to the same design with FP16 KV
caches (W2KV16).}
\label{fig:eval-kv}
\end{figure}
% [modeled] evaluation/data/sensitivity.csv, kv_contribution.csv.
% FFN T/T_FP16 b1..b32: VQACC AQLM-1 0.424--0.468, GPTVQ-2 0.452--0.534;
% EVA-OL32 0.063/0.145 at b1, >1 from b16 (GPTVQ-2) / b32 (AQLM-1).
% KV (vs own FP16 KV): VQACC CQ-2 0.931 (1K) / 0.748 (32K); VPU CQ-2 1.019--1.085,
% CQ-4 1.029--1.135.

Fig.~\ref{fig:eval-batch} details the effect of the batch size across designs:
at every batch size, VQACC is faster than every baseline except
EVA-OL32. In contrast, EVA-OL32 is the best only up to batch 8, where the
reduced computation of its transformation pays off, but as the batch grows its
latency grows with the computation. This limitation restricts its usability in
the mainstream multi-batch decoding setting, and rules out prefill, where the
computation is far larger than in decoding.
Fig.~\ref{fig:eval-kv} further shows the effect of the KV cache length: as the
length grows, the advantage of VQACC and Vortex-IL becomes more pronounced but
approaches a plateau (for KV2, the latency ratio levels off at 0.75$\times$ at
32K tokens), because the codebook-sharing partition specific to
attention caps the compute tile at a small size, so on-chip compute time is
dominated by SA drain cycles and becomes the bottleneck. 

\subsection{Prefill Performance}
\label{sec:eval-prefill}
\begin{figure}[!t]
\centering
\includegraphics[width=0.95\columnwidth]{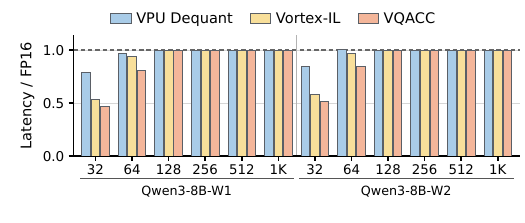}
\caption{Prefill latency of Qwen3-8B at batch 1 with context lengths 32--1K,
normalized to Full FP16.}
\label{fig:eval-prefill}
\end{figure}
% [modeled] evaluation/data/prefill.csv. Reduction: VQACC 53.4%/48.9% at S32
% (AQLM-1/GPTVQ-2); S32 Vortex-IL 46.4/42.0%, VPU 21.4/15.0%; S64 VQACC 19.1/15.1%,
% Vortex-IL 5.7/3.6%, VPU 3.1/-0.5%; S>=128 (M128 K128 N128 tiles, SA-bound)
% VQACC and Vortex-IL 0.63--0.66%, VPU 0.08--0.14%.

Although prefill is not the target of VQ, running the quantized weights
there is still worth evaluating (Sec.~\ref{sec:bg-quant}).
Fig.~\ref{fig:eval-prefill} shows that VQACC's prefill is at least as fast as
Full FP16 at every context length, and sometimes significantly faster. This advantage comes
from the DRAM traffic that VQ itself removes, so it is clearest on short
contexts, reaching 53\% at 32 tokens. As the context grows, the bottleneck
shifts gradually to computation, and latency levels off with Full FP16.

\subsection{Area and Energy}
\label{sec:area}
\label{sec:energy}
\begin{figure}[t]
\centering
\includegraphics[width=0.95\columnwidth]{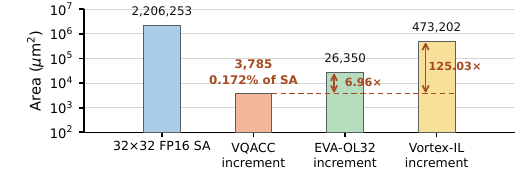}
\caption{Area of the SA and the components added by VQACC, EVA-OL32 (the
additional reduction tree for output lookup), and Vortex-IL (codebook and
staging buffers, lookup logic) in 28\,nm (log scale).}
\label{fig:eval-area}
\end{figure}
\begin{figure}[!t]
\centering
\includegraphics[width=0.95\columnwidth]{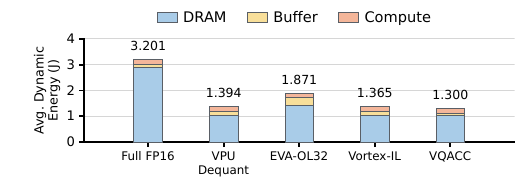}
\caption{Average energy per decoding step over the workloads in
Fig.~\ref{fig:eval-overall}, broken down into DRAM, buffer, and compute.}
\label{fig:eval-energy}
\end{figure}
% [reported] notes/area-comparison.json: SA 2206252.600816 um^2; VQACC
% shift network (32-lane x 16-bit rotator, five 2:1 mux stages, combinational,
% 1 GHz setup slack +0.003826 ns, no pipeline register;
% rtl_fp16/evaluation/20261118_rotate32) 2747.135941 + index address
% generator 1037.568003 = 3784.703944 um^2 = 0.17154% of SA;
% EVA-OL32 reduction tree 26350.463761 um^2 = 6.96x; Vortex-IL 2x16 KiB codebook
% buffer 310407.66 + 2x8 KiB staging buffer 159009.72 um^2 (author
% memory-compiler measurements) + lookup logic charged as VQACC's 3784.703944
% = 473202.083944 um^2 = 125.03x VQACC. Component areas, not total area.
% [modeled] evaluation/data/energy_summary.json; coefficients in
% notes/energy-model.json. Dynamic energy only. DRAM/buffer J: VQACC
% 1.033/0.077, Vortex-IL 1.033/0.142, VPU 1.033/0.171, EVA-OL32 1.419/0.299;
% total VPU Dequant. 1.394 J.

VQACC's gains cost almost no area. RTL synthesis shows that the logic VQACC
adds occupies only 3,785\,$\mu$m$^2$, just 0.17\% of the SA
(Fig.~\ref{fig:eval-area}). By comparison, the additional area of EVA-OL32 and
Vortex-IL is 7.0$\times$ and 125.0$\times$ larger, respectively. Since Vortex
does not disclose its microarchitecture details, we count only the cost of
its additional buffers, so its actual overhead is larger. Moreover, we evaluate the energy gap in decoding: VQACC achieves
the lowest dynamic energy per decoding step among all designs
(Fig.~\ref{fig:eval-energy}).

\subsection{Codebook-Stationary Dataflow}
\label{sec:eval-dataflow}
\begin{figure}[!t]
\centering
\includegraphics[width=\columnwidth]{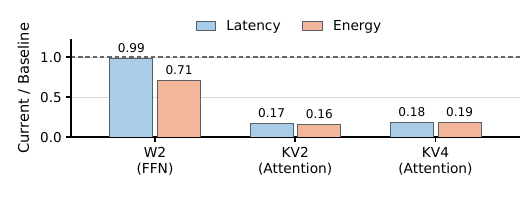}
\caption{Latency and energy of the codebook-stationary dataflow relative to
a codebook-unaware baseline (Qwen3-8B; mean over batch 1--32).}
\label{fig:eval-dataflow}
\end{figure}
% [modeled] scripts/codebook_dataflow_ablation.py -> evaluation/data/
% codebook_dataflow.csv; plot_dataflow_ablation.py. Current/baseline over b1-32:
% W2 latency 0.991 (0.967--0.998), energy 0.707 (0.578 at b1 -- 0.790 at b32);
% KV2 0.174/0.164; KV4 0.185/0.187; codebook loads 8x (W2) / 16x (KV) fewer.

Fig.~\ref{fig:eval-dataflow} compares the conventional and codebook-stationary dataflows on VQACC. In the attention computation, the
conventional dataflow switches codebooks frequently, and the small SA compute
tiles (Sec.~\ref{sec:eval-overall}) cannot hide that latency, so it incurs
5.4--5.8$\times$ the latency and 5.3--6.1$\times$ the energy of the
codebook-stationary dataflow. In the FFN, W2's large $C_K$ and $C_N$ give large
tiles that hide the switches, so the latency gain is small, yet 8$\times$ fewer
codebook loads still lower FFN energy by 29\% on average.

\section{Conclusion}
\label{sec:conclusion}

Vector quantization is a key enabler of ultra-low-bit LLM inference, yet its
dequantization flow is a significant performance bottleneck. This paper
presented VQACC, which unlocks efficient VQ inference on existing SA-based LLM
accelerators. VQACC adds no separate dequantization stage;
instead, it reuses the storage, ports, and addressing of the existing weight
buffer, so that codebook lookups are folded into ordinary operand reads. It achieves overall
speedups of 2.63$\times$ and 1.10$\times$ over the existing VQ LLM accelerators
EVA and Vortex, and occupies only 0.17\% of the compute array's area.

% arXiv version: no references-only page; the bibliography follows the
% conclusion directly.
\label{endofbody}
\bibliographystyle{IEEEtranS}
\bibliography{refs}

\end{document}